\documentstyle[prl,aps,multicol,epsf]{revtex}

\begin{document}
\begin{multicols}{2}
\narrowtext

\noindent{\bf Pagnani, Parisi and Ricci-Tersenghi reply:} In the
preceding Comment~\cite{COMMENT} to our paper~\cite{US} Hartmann
presents a powerful algorithm to find the ground states of the random
RNA model we have studied in~\cite{US}.  He also shows some
interesting results on the overlap distribution $P(q)$ at zero
temperature.  His conclusion is that at $T=0$ and in the
thermodynamical limit the $P(q)$ is a delta function.  We would like
to point out that his result is not in contradiction with ours and
that none of the conclusions we reached in~\cite{US} is to be
modified.

In our paper we already commented about the shrinking of the $P(q)$ at
$T=0$, finding that the variance decreases like $\sigma^2 \propto
L^{-0.4}$.  However, because of the small exponent, we were not able
to determine the asymptotic value of the variance from our data ($L
\le 1024$) and we believe that even with Hartmann's data ($L \le
2000$) this extrapolation is still a hard task.  Note that the
relevant observable is the width $\sigma$ which scales with an
exponent of order $0.2 \div 0.25$, according to ours or Hartmann
results.  We also measured~\cite{US} the width in the high temperature
phase and we obtained $\sigma \propto L^{-0.5}$ as it should be
according to the central limit theorem.

Nevertheless, assuming that according to~\cite{COMMENT} the model
studied in~\cite{US} at $T=0$ has a delta-shaped $P(q)$, our
conclusions regarding the replica symmetry breaking (RSB) transition
will remain unaltered.  Indeed there are disordered models which have
an RSB phase together with a trivial $P(q)$ at $T=0$, showing that the
two properties are completely unrelated.  The most famous among these
is the Sherrington-Kirkpatrick (SK) one~\cite{SK}, which is widely
considered the prototype of disordered models.  Below the critical
temperature the SK model has an RSB phase~\cite{MPV}, however at $T=0$
the $P(q)$ is a delta function centered in $q=1$.

An expert reader may object that, differently from the SK model, the
random RNA model we have studied in~\cite{US} has a finite entropy at
$T=0$ and then the above argument may no longer hold.  However we do
not expect any significant difference and we corroborate our belief
with the most recent results on the Edwards-Anderson (EA) model with
discrete couplings ($\pm$J) in dimensions $d=3,4$.  In this model at
$T=0$ there is a finite entropy and a trivial delta-shaped
$P(q)$~\cite{HARTMANN}, however as soon as the temperature is
different from zero the $P(q)$ becomes broad~\cite{GS}.  Moreover in
the $\pm$J EA model, independently from the existence of a finite
temperature phase transition (which is present in $d=3,4$ and absent
in $d=2$), the $P(q)$ at $T=0$ is always a delta
function~\cite{HARTMANN}.  If one is interested in understanding the
presence of a finite temperature phase transition through the ground
states calculation, methods more sophisticated than the simple $P(q)$
estimation should be used~\cite{GS}.  Their application to the random
RNA models studied in~\cite{US,HIGGS} would be very welcome.

The above observations should clarify the importance of studying the
model at finite temperatures, where a broad $P(q)$ indeed signals the
presence of an RSB phase.  We presented the key results obtained at
$T\neq0$ in the third figure of~\cite{US}.  They show the existence of
a phase transition to a phase where the replica symmetry seems to be
broken.  Moreover in the discussion following Fig.~3 in~\cite{US} we
also took into account the possibility that the RSB would be simply
given by finite size effects and that it would disappear in the
thermodynamical limit.  Even in this case the zero-energy excitations
between different ``valleys'', which mimic the presence of RSB, would
play a fundamental role in the understanding of the model.

Finally, in order to confirm the glassy nature of the transition at
finite temperature, we are currently implementing a Monte Carlo study
of the model dynamics.  Preliminary results clearly display typical
glassy features, like aging.  We hope that these results will
eventually render a coherent scenario of RSB both from dynamical and
thermodynamical point of view.

\vspace{1cm}
\noindent A. Pagnani$^a$, G. Parisi$^b$ and F. Ricci-Tersenghi$^c$
\vspace{0.5cm}

\noindent{\small $^a$ Dipartimento di Fisica and INFM,
Universit\`a di Roma {\it Tor Vergata},
Via della Ricerca Scientifica 1, 00133 Roma (Italy)\\
$^b$ Dipartimento di Fisica and INFN,
Universit\`a di Roma ``La Sapienza'',
Piazzale Aldo Moro 2, 00185 Roma (Italy)\\
$^c$ Abdus Salam ICTP, Condensed Matter Group,
Strada Costiera 11, P.O. Box 586, 34100 Trieste (Italy)}

\end{multicols}
\end{document}